\documentstyle[11pt,cite,epsfig,rotating]{article}
\newcommand{\mr}[1]{{\mathrm {#1}}}

\newcommand{\lm}{$\Lambda\;$}
\newcommand{\lme}{\Lambda}

\newcommand{\lmepdf}{\Lambda^{\mr{pdf}}\;}

\newcommand{\lmehsc}{\Lambda^{\mr{hsc}}\;}
\newcommand{\lms}{$\Lambda_{\overline{\mr{MS}}}\;$}
\newcommand{\lmse}{\Lambda_{\overline{\mr{MS}}}}

\newcommand{\bbare}[1]{\overline{\mr{#1}}}
\newcommand{\bbar}[1]{$\overline{\mr{#1}}$}
\newcommand{\reff}[1]{(\ref{#1})}
\newcommand{\be}{\begin{equation}}
\newcommand{\ee}{\end{equation}}
\newcommand{\bear}{\begin{eqnarray}}
\newcommand{\eear}{\end{eqnarray}}
\newcommand{\bc}{\begin{center}}
\newcommand{\ec}{\end{center}}
\newcommand{\bd}{\begin{description}}
\newcommand{\ed}{\end{description}}
\newcommand{\bit}{\begin{itemize}}
\newcommand{\eit}{\end{itemize}}
\newcommand{\ben}{\begin{enumerate}}
\newcommand{\een}{\end{enumerate}}

\newcommand{\als}{$\alpha_s(M_Z)\;$}

\begin{document}
\begin{flushright}
PRA-HEP/96-01
\end{flushright}
\bc
\vspace*{0.5cm}
\noindent
{\huge \bf On consistent  determination of $\alpha_s$
from jet rates in deep inelastic scattering}\\
\vspace*{0.7cm}
{\large Ji\v{r}\'{\i} Ch\'{y}la and Ji\v{r}\'{\i} Rame\v{s}} \\
\vspace*{0.4cm}
Institute of Physics, Academy of Sciences of the Czech Republic, Prague
\footnote{e--mail: chyla@fzu.cz, rames@fzu.cz}

\vspace*{1cm}
Abstract \\
\ec

\noindent
For theoretically consistent
determination of $\alpha_s$ from jet rates in deep inelastic scattering
the dependence on $\alpha_s$ of parton distribution functions is in
principle as important as that of hard scattering cross--sections.
For the kinematical region accessible at HERA we investigate in detail
numerical importance of these two sources of the $\alpha_s$
dependence of jet rates.

\section{Introduction}
As QCD enters the stage of a truly quantitative theory, great emphasize
is put on precise determination of its basic parameter, $\alpha_s$, or
equivalently \lms, using different sorts of hard scattering processes in
broad range of kinematical conditions. As a result of numerous
phenomenological analyses crucial property of QCD -- the running of
$\alpha_s$ -- has been verified beyond doubt (for review see
\cite{Bethke,Bryan}). However, the increasing precision of
these measurements has lead to a discrepancy between \als
extrapolated from measurements of $\alpha_s(\mu)$ at low scales $\mu$,
and \als measured directly at LEP. This effect, if confirmed by more
accurate data, would indicate that the running of $\alpha_s$ slows
down in the LEP range.  This, in turn, might signal the long sought new
physics \cite{Schifman}.

One of the problems of quantitative determination of the running of
$\alpha_s$ is related to the fact that it usually requires combining
results of different experiments in different kinematical regions.
One of the hard scattering processes where the running of
$\alpha_s(\mu)$ can be investigated for the broad range of scales
in a single experiment is the deep inelastic
scattering (DIS) at HERA. The theoretically cleanest way for such
an investigation, the measurement of scaling violations of
$F_2^{\mr{ep}}(x,Q^2)$, is marred by the fact that scaling
violations are tiny effect at large $Q^2$.  Recently, however, both the
H1 \cite{H1} and ZEUS \cite{ZEUS} Collaborations have reported
evidence for the running of $\alpha_s(\mu)$ obtained from the
measurement of jet rates in DIS via the quantity
\be
R_{2+1}(Q^2)\equiv \frac{\sigma_{2+1}(Q^2)}
{\sigma_{1+1}(Q^2)+\sigma_{2+1}(Q^2)},
\label{R2+1}
\ee
where $\sigma_{k+1}$ denotes the cross--section for the production
of $k$ hard and one proton remnant jets. There are two
Monte--Carlo implementations \cite{PROJET,MEPJET} of the NLO
calculations \cite{G1,G2,B1,B2} of the cross--sections
$\sigma_{k+1}(Q^2)$, which include also the possibility of imposing cuts
on various kinematical variables. In this paper we shall concentrate on
the H1 analysis \cite{H1} of 1993 data, which, using PROJET 4.1
\cite{PROJET} with the JADE
jet algorithm and $y_c=0.02$, obtained the following result for
$\alpha_s(M_Z,\bbare{MS})$
\be
\alpha_s(M_Z,\bbare{MS})=
0.123\pm0.012(\mr{stat.})\pm0.008(\mr{syst.}).
\label{alphas}
\ee
This value has been
extracted from the $Q^2$--dependence of $R_{2+1}(Q^2)$, displayed in
Fig. 5 of \cite{H1}, using only the two highest $Q^2$ data points. 
Within the error bars the value \reff{alphas} is consistent with the 
world average, but its accuracy is insufficient to draw any firmer 
conclusions.  Discarding (for reasons discussed in \cite{H1}) the three 
lowest $Q^2$ points means not taking into account the region where most 
of the running of $\alpha_s(\mu)$ occurs and where also the data are 
most sensitive to \lms. The experimental errors would have to decrease 
by at least a factor of three to make this way of determining
$\alpha_s(M_Z,\bbare{MS})$ competitive with the most precise
determinations of $\alpha_s(M_Z,\bbare{MS})$ based on
DIS scattering or LEP event shapes data. Some improvement may be
expected already from the ongoing analyses of 1994 data.

There is one aspect of the extraction of $\alpha_s$ from the measured
ratio \reff{R2+1} crucial for its consistency. In \cite{H1}
$\alpha_s(\mu,\bbare{MS})$
(or, equivalently, $\lmse$) was considered as a
free parameter in the hard scattering cross--sections, but not in the
parton distribution functions (PDF), for which the MRSH set was used.
The authors of \cite{H1} concede that ``for strict consistency'' one
should ``reevolve the distributions'' using the value of
$\alpha_s(\mu,\bbare{MS})$ obtained in their paper, but
claim that given the presently obtainable accuracy ``this is not
necessary''. However, reevolving PDF, that is solving the evolution
equations with the initial conditions corresponding to the MRSH (or any
other chosen set of PDF), but using the value of \lms extracted from
\reff{R2+1} instead of the ``input'' $\lmse(\mr{MRSH})$ would solve
nothing. Except for the case that the extracted \lms coincided with the
input one, the reevolved PDF would be different from the original
MRSH ones. This would require iterating the whole procedure until the
extracted value of \lms coincided with the input one. But then one would
have to check whether the resulting new PDF still described the whole
body of hard scattering data on which the original global fit was based.
Moreover, although mathematically legal exercise, reevolving PDF from
unchanged initial conditions with changed \lms is, as argued in Section
3 below, theoretically inconsistent. In the situation when  the
measurement of the quantity \reff{R2+1} alone is clearly insufficient to
determine $\alpha_s$, the only meaningful way of using its results is to
include them in the conventional global fits and see what comes out.
\footnote{This is what the MRS group did in their latest global fit
\cite{GMRS}, which includes beyond the standard set of hard scattering
data also the very recent CDF results on jet production \cite{CDF}.}

Nevertheless, it might be useful to know under which conditions
neglecting \lms in the PDF is {\em numerically} good approximation
for determining $\alpha_s$ from \reff{R2+1}. The
purpose of our paper is to investigate this question in detail.
The paper is organized as follows. In the next section we recall basic
relevant formulae and emphasize the importance of considering \lms
in PDF. In Section 3 a simple procedure of
constructing PDF for arbitrary \lms is outlined. Results of numerical
simulations in which \lms is varied separately in hard scattering
cross--sections and PDF are presented and discussed in Section 4,
followed by conclusions in Section 5. Detailed elaboration of
arguments supporting the claim made in Section 2 is left for the
Appendix.

\section{Basic formulae}
We use
the couplant $a\equiv\alpha_s(\mu)/\pi$ as QCD expansion parameter.
It satisfies the equation
\be
\frac{\mr{d}a(\mu,\mr{RS})}{\mr{d}\ln \mu}=
-ba^2(\mu,\mr{RS})\left(1+ca(\mu,\mr{RS})
+c_2a^2(\mu,\mr{RS})+\cdots\right).
\label{RG}
\ee
For 3 colours and $n_f$ massless quark flavours
$b=(33-2n_f)/6,\;c=(153-19n_f)/(66-4n_f)$,
while all higher order coefficients $c_k;k\ge 2$ in \reff{RG} are
free parameters, which define 
the so called renormalization convention (RC).
Together with the specification of the initial condition on the solution
of \reff{RG} they define the renormalization scheme (RS). To get a unique
result of perturbative calculations both the scale $\mu$ and the RS must
be specified. Throughout
this paper we stay in the conventional \bbar{MS} RS and therefore
the corresponding specification ``\bbar{MS}'' in \lms as well as in
$\alpha_s(\mu,\bbare{MS})$ will be dropped in the following.
The jet cross--sections $\sigma_{k+1}$ in \reff{R2+1} are given as
integrals over the parton level cross--sections $C_{k+1,i}$
\be
\sigma_{k+1}(Q^2,y_c,\lme)\equiv \sum_{i}\int^1_0 \mr{d}x
f_{i}(x,M,\lme)C_{k+1,i}(Q,M,x,y_c,\lme),
\label{sigmak+1}
\ee
where $f_i(x,M)$ are PDF of the
proton evaluated at the factorization scale $M$ and the sum runs over
all parton species, i.e. quark, antiquarks and gluon.
In \reff{sigmak+1} we have written out explicitly
the dependence of both jet and parton level hard scattering
cross--sections $\sigma_{k+1},C_{k+1,i}$ as well as of the PDF
$f_i$ on \lm. This dependence reflects the fact that
physical quantities do depend on the value of \lm (in a chosen
RS). Contrary to that, jet cross-sections $\sigma_{k+1}$ evaluated to 
all orders of perturbation theory are independent of both the 
factorization scale $M$ and the RS. In perturbative QCD the parton level 
hard scattering cross-sections are given as expansions in the couplant 
$a(\mu/\lme)$, taken at the hard scattering scattering scale 
\footnote{Although in general $\mu\neq M$, we shall in the rest of this 
paper follow the usual practice and set $\mu=M$.} $\mu$
\begin{eqnarray}
C_{2+1,i}(Q,M,x,y_c,\lme)& = &
a(\mu/\lme)\left[c^{(1)}_{2+1,i}(x,y_c)+
a(\mu/\lme)c^{(2)}_{2+1,i}(Q/M,x,y_c)\right],
\label{C2+1}    \\
C_{1+1,i}(Q,M,x,y_c,\lme) & = &
c^{(0)}_{1+1,i}(Q,x)+
 a(\mu/\lme)c^{(1)}_{1+1,i}(Q/M,x,y_c).
\label{C1+1}
\end{eqnarray}
In standard global analyses of hard scattering processes
\cite{MRS,CTEQ,GRV},
\lm is fitted together with a set of parameters
$a_i^{(j)}$, describing distribution functions $p^{(j)}(x,M)$
of parton species $j$ at some initial scale $M_0$, usually in the
form
\be
p^{(j)}(x,M_0)=a_0^{(j)}
x^{a_1^{(j)}}(1-x)^{a_2^{(j)}}
\left(1+a_3^{(j)}\sqrt{x}+a_4^{(j)}x+a_5^{(j)}x^{3/2}\right).
\label{initial}
\ee
To determine the relative importance of  varying \lm in PDF and hard
scattering cross--sections, we write the derivative
$\mr{d}\sigma_{2+1}/\mr{d}\ln \lme$:
\begin{eqnarray}
\lefteqn{\frac{\mr{d}\sigma_{2+1}(Q^2,\lme)}{\mr{d}\ln\Lambda}=
\sum_{i}\int^1_0 \mr{d}x\left[
\frac{\mr{d}f_{i}(x,M,\lme)}{\mr{d}\ln\Lambda}
a(M)c_{2+1,i}^{(1)}(x)+f_i(x,M,\lme)c_{2+1,i}^{(1)}(x)
\frac{\mr{d}a(M)}{\mr{d}\ln\lme}\right]=}  \label{l1} \nonumber \\
\!\!\! & \! & a(M)\sum_{ij}\int^1_0 \mr{d}x\left[-
\int^1_0\frac{\mr{d}y}{y}f_j(y,M,\lme)P^{(0)}_{ij}(z)
a(M)c_{2+1,i}^{(1)}(x)+bf_i(x,M,\lme)
a(M)c_{2+1,i}^{(1)}(x)\right], \label{l3}
\end{eqnarray}
where $z\equiv x/y$ and $P^{(0)}_{ij}(z)$ are the LO AP branching
functions
\footnote{In (8) we have for brevity suppressed
the dependence of $c^{(1)}_{2+1,i}$ on $y_c$ and written $a(M/\lme)$
simply as $a(M)$.}.
In the above expression we have used \reff{RG} and standard
evolution equations for PDF, exploited the equality
$\mr{d}a(M/\lme)/\mr{d}\ln M= -\mr{d}a(M/\lme)/\mr{d}\ln\lme$
and took into account that the same relation between
derivatives with respect to $M$ and $\lme$ holds also for PDF.
The first term in the brackets of \reff{l3}
results from variation of \lm in PDF while the second from variation of
\lm in parton level hard scattering cross--sections.
Eq. \reff{l3} implies that the leading order term of
$\mr{d}\sigma_{2+1}/\mr{d}\ln \lme$ involves merely the
LO evolution equations for PDF and leading order term of parton level
hard scattering cross--sections $C_{2+1,i}$. The dependence
of $\sigma_{2+1}$ on \lm starts therefore already at the LO and
consequently also the relative importance of the two terms in the
brackets of \reff{l3} is basically the LO effect. Note, however, that
associating \lm in $\sigma_{2+1}$ with any well-defined RS requires
including the NLO corrections as well.
Eq. \reff{l3} tells us that the variation of \lm in the couplant
$a(\mu/\lme)$ changes the jet cross--section $\sigma_{2+1}$ by the
term of the same order (schematically $O(a)\sigma_{2+1}$) as when \lm 
is varied in PDF. Consequently if we want to determine
\lm from jet cross--section $\sigma_{2+1}$, we {\em must} vary it in PDF
as well.

For the quantity $R_{2+1}(Q^2)$ the situation is less obvious. PDF
appear in both the numerator and denominator of \reff{R2+1} and so some
cancellations might occur, while the hard scattering cross--sections
start as $O(a)$ for $C_{2+1,i}$ and as $O(1)$ for $C_{1+1,i}$.
Nevertheless as shown in the Appendix also for the ratio
$R_{2+1}$ varying \lm in PDF generates terms which are of the same order
as those resulting from the variation of \lm in the hard scattering
cross-sections $C_{k+1,i}$.

\section{Parton distribution functions for arbitrary \lm}
In the previous Section we have shown that for a theoretically
consistent extraction of \lm from $R_{2+1}(Q^2)$, \lm must be
varied not only in hard scattering cross--sections, but also in the PDF.
We now want to investigate the numerical importance of varying \lm
in PDF. For this we first have to know what to do with the parameters
$M_0,a^{(j)}_i$, specifying the initial conditions \reff{initial},
when \lm is varied. The obvious choice seems to be to keep them fixed
and simply solve the evolution equations with variable \lm.
This, however, would be inconsistent with the idea of factorization,
which implies that also the initial conditions depend on $\Lambda$!

To see the caveat, let us first consider \reff{RG} as well as the
evolution equation for the nonsinglet quark distribution function
$q_{\mr{NS}}(x,M)$ at the LO. In term of moments, defined as
\be
f(n)\equiv \int^1_0 \mr{d}x x^n f(x)
\label{definicemom}
\ee
we have
\be
\frac{\mr{d}q_{\mr{NS}}(n,M,\lme)}{\mr{d}\ln M}=a(M/\lme)
q_{\mr{NS}}(n,M)P^{(0)}_{\mr{NS}}(n),\;\;
d_n\equiv -P^{(0)}_{\mr{NS}}(n)/b,
\label{LLmom}
\ee
wherefrom
\be
q_{\mr{NS}}(n,M,\lme)=
A_{n}\left[a(M/\lme)\right]^{d_n},
\;\;\;a(M/\lme)=\frac{1}{b\ln (M/\lme)}.
\label{LLsolution}
\ee
In the above expressions
$P^{(0)}_{\mr{NS}}(n,M)$ are moments of LO nonsinglet AP branching
function and $A_n$ are dimensionless constants describing the
nonperturbative, long distance, properties of the nucleon.
These constants are unique pure numbers.
Via \reff{LLsolution} they determine the asymptotic
behavior of $q_{\mr{NS}}(n,M,\lme)$ for $M\rightarrow \infty$
\cite{Politzer} and thus provide alternative way of specifying the
initial conditions on the solutions of \reff{LLmom}.
The separation in \reff{LLsolution} of $q_{\mr{NS}}(n,M,\lme)$ into two
parts -- one calculable in perturbation theory and the other
incorporating all the nonperturbative effects -- is the essence of the
factorization idea \cite{Politzer}. Note that in
\reff{LLsolution} the dependence on \lm is simply a reflection of its
dependence on $M$. Knowing the latter, we know the former.  For finite
initial $M_0$ \reff{LLsolution} implies
\be
q_{\mr{NS}}(n,M,\lme)=q_{\mr{NS}}(n,M_0,\lme)
\left[\frac{a(M/\lme)}{a(M_0/\lme)}\right]^{d_n}=
q_{\mr{NS}}(n,M_0,\lme)\exp(-d_n s),
\label{MM0}
\ee
where
\be
s\equiv\ln\left(\frac{\ln(M/\lme)}{\ln(M_0/\lme)}\right)
\doteq
\ln\left(\frac{a(M_0/\lme)}{a(M/\lme)}\right)
\label{s}
\ee
is the so called ``evolution distance'' \cite{GRV}. The second equality
in \reff{s} holds exactly at the leading order and approximately at
higher ones. Rewriting \reff{MM0} again in the form \reff{LLsolution}
\be
q_{\mr{NS}}(n,M,\lme)=\left[\frac{q_{\mr{NS}}(n,M_0,\lme)}
{\left(a(M_0/\lme)\right)^{d_n}}\right]
\left(a(M/\lme)\right)^{d_n}
\label{ratio}
\ee
we see that the ratio of $q_{\mr{NS}}(n,M_0,\lme)$ and
$(a(M_0/\lme))^{d_n}$ equals the invariant $A_n$ and thus is both
$M_0$ and \lm independent. This, however, is possible only if the
initial moments $q_{\mr{NS}}(n,M_0,\lme)$ do, as already indicated,
depend on \lm as well! What happens in \reff{ratio} is that the \lm
and $M_0$ dependencies coming from $a(M_0/\lme)$ in the brackets of
\reff{MM0} cancel those of the initial $q_{\mr{NS}}(n,\lme,M_0)$.
Note that considering $q_{\mr{NS}}(n,M,\lme)$ as a
function of $s$, formula \reff{ratio} implies that entire 
dependence of $q_{\mr{NS}}(n,M,\lme)$ on \lm
comes solely from the term $\ln(M/\lme)=1/a(M/\lme)$ in \reff{s}! The
same happens in the realistic case of coupled quark and gluon evolution
equations \cite{Buras}. The dependence of PDF on $M$ and $\lme$ through
the evolution distance variable $s$ is then a direct consequence of the
above relations for the moments.

To investigate the dependence of PDF on \lm we
therefore cannot fix initial conditions at some finite $M_0$, specified
by  $p^{(j)}(x,M_0)$ in \reff{initial}, and vary \lm in the evolution
equations alone. The preceding discussion also suggests what can be held
fixed in the process of varying \lm -- the constants $A_n$!
There is no principal obstacle to carrying out QCD analyses with the
initial conditions specified by means of the invariants $A_n$, which
together with \lm should then be fitted to data. There is also a well 
developed technique of orthogonal polynomials expansions \cite{Barker}, 
which could be used for this purpose. There are, however, technical 
problems in practical implementation of this method. One of them 
concerns the number of parameters $A_n$ required for a specification of 
initial conditions. As shown in \cite{my}, one needs at least 9 to 10 
first moments of PDF to achieve sufficient precision in reconstruction 
of PDF from their moments. That is about twice the number of free 
parameters used in conventional initial conditions \reff{initial} at 
finite $M_0$ and seems too much for any reasonable QCD fit.

However, the preceding paragraph also suggests a simple procedure,
which makes use of some of the available parameterizations and
generalizes them to arbitrary \lm, keeping, as discussed above, the
constants $A_n$ fixed. From all the available parameterizations of PDF 
those given by analytic expressions of the coefficients $a^{(j)}_i$ on a 
general factorization scale $M$ via the variable $s$, defined in 
\reff{s}, are particularly suitable for this purpose. In our studies we 
took several of the GRV parameterizations, both LO and NLO ones, defined 
in \cite{GRV2}. At the LO the recipe for  construction of PDF for 
arbitrary \lm is simple:
\bit
\item Use any of the parameterizations expressing the solutions of
evolution equations as functions of the ``evolution distance'' variable
s, defined in \reff{s}.
\item Keep $\lme$ in $\ln(M_0/\lme)$ fixed at the value obtained
in the fit.
\item Vary $\lme$ in $\ln(M/\lme)$.
\item Use the original parameterization with variable $s=s(\lme)$.
\eit
This construction is exact at the LO. At the NLO the couplant
$a(M/\lme)$ is no longer given simply as $1/\ln(M/\lme)$ and
consequently  the second equality in \reff{s} holds only approximately.
This, together with the additional term appearing in the NLO expression
for $q_{\mr{NS}}(n,M,\lme)$
\be
q_{\mr{NS}}(n,M,\lme,A_n)=
A_{n}\left[\frac{a(M/\lme)}
{1+ca(M/\lme)}\right]^{d_n}
\left(1+ca(M/\lme)\right)^{P^{(1)}_{\mr{NS}}(n)/bc},
\label{NLLsolution}
\ee
destroy simple dependence of $q_{\mr{NS}}$ on $M,M_0$ and $\lme$ via the
variable $s$. Using the above recipe for the NLO GRV parameterizations
provides therefore merely an approximate solution of our task, but one
that is sufficient for the purposes of determining the relative
importance of varying \lm in PDF and hard scattering cross--sections.

As an illustration of our procedure we plot in Fig. 1, for $x\ge
10^{-4}$ and two values of $Q^2$, the dependence of valence and sea
$u$--quark and gluon distribution functions on $\lme$. The curves
correspond to the LO GRV parameterization of \cite{GRV2}.
We see that this
dependence decreases with increasing $Q^2$, is most pronounced for the
gluon distribution function and almost irrelevant for the valence quark
one. These features are qualitatively the same as those obtained in
refs. \cite{Andreas,Alan}, which contain results of global fits
performed for several fixed values of $\lme$. Nevertheless,
quantitatively the sensitivity to the variation of \lm, observed in
these papers, is markedly weaker than that of our results, displayed in
Fig. 1.  The reason for this difference is simple. In
global analyses of \cite{Andreas,Alan} fitting the data for different
values of $\lme$ leads to different values of boundary condition
parameters $a^{(j)}_i$, which partially compensates the variation of
$\lme$. Eqs. \reff{LLsolution} or \reff{NLLsolution} tell us that for
any fixed scale $M$ and for any moment $n$ (and thus for any $x$) the
change of $\lme$ can be {\em fully} compensated by an appropriate change
of the constants $A_n$.
For a finite interval of $M$ this compensation cannot be complete, but 
it is clear that the dependence of the fitted PDF on $\lme$ comes out 
weaker than in our procedure where only $\lme$ is varied and all the 
coefficients $A_n$ (or, analogously, $a^{(j)}_i$) are held fixed.

\section{Numerical results}
All the results reported below were obtained using the PROJET 4.1
generator \cite{PROJET}. To quantify the dependence of $R_{2+1}$ on \lm
separately in hard scattering cross--sections and PDF we consider it
as a function of $Q^2, y_c$ and two independent $\Lambda$--parameters,
denoted as $\lme^{\mr{hsc}}$ and $\lme^{\mr{pdf}}$ respectively. The
results are then studied subsequently as a function of $\lmehsc,
\lmepdf, Q^2$ and $y_c$.  First we analyse the case with no
cuts on the final state jets and then discuss the changes caused by the
imposition of cuts as specified in \cite{H1}.

\subsection{The case of no cuts}
The simulations were done at both the leading and next--to--leading
order and for
\bit
\item 12 values of $Q^2$ (equidistant in $\ln Q^2$) between 5 and
$10^4$ GeV$^2$,
\item 5 values of $\lmepdf$ or $\lmehsc$, equal to $0.1,0.2,0.3,0.4$
and $0.5$ GeV,
\item 3 values of $y_c=0.01,0.02,0.04$.
\eit
First we discuss the results of the NLO analysis.
Fig. 2 shows a series of plots corresponding to $y_c=0.01$ and
displaying the dependence of $R_{2+1}$ on $\lmehsc$ for fixed $\lmepdf$
(dotted curves) or on $\lmepdf$ for fixed $\lmehsc$ (solid curves).
Each plot corresponds to one of the 6 selected values of $Q^2$. 
We see that for $Q^2$ below about $40$ GeV$^2$ the solid curves are
steeper that the dotted ones, while above $40$ GeV$^2$ the situation is
reversed. This means that below $40$ GeV$^2$ varying \lm in PDF is more
important than varying it in the hard scattering cross--sections for
$Q^2$, while the opposite holds above $40$ GeV$^2$. The same
information as in Fig. 2 is displayed in a different manner
in Fig. 3, where the $Q^2$ dependence of $R_{2+1}$ is plotted for
various combinations of $\lmepdf$ and $\lmehsc$. The shape of the curves
in Fig. 3 is a result of two opposite effects: as $Q^2$ increases the
phase space available for produced jets increases as well, but at the
same time the value of the running $\alpha_s(Q/\lme)$ decreases. At low
$Q^2$, on the other hand, phase space shrinks, but $\alpha_s(Q/\Lambda)$
grows. The same features of the relative importance
of varying $\lmepdf$ and $\lmehsc$ as mentioned above are reflected in a
bigger spread of the five curves in Figs. 3d)-f) compared to those in
Fig. 3a)-c) for $Q^2\le 40$ GeV$^2$ and smaller spread above $40$
GeV$^2$. Note also that for fixed $\lmepdf$ the sensitivity to
the variation of $\lmehsc$ is about the same at all $Q^2$. The shape of
curves in Fig. 2 suggests that the ratio
\be
V(Q^2,y_c,r,s)\equiv
     \frac{R_{2+1}(Q^2,y_c,\lmepdf=s,\lmehsc=r)}
{R_{2+1}(Q^2,y_c,\lmepdf=r,\lmehsc=s)},
\label{V}
\ee
is to a good approximation a linear function of $s$ for any fixed
$Q^2,y_c,r$. Considered as a function of $s$ for fixed $r$,
\reff{V} quantifies the relative importance of varying \lm in the
PDF (described by $R_{2+1}$ in the numerator) and in hard scattering
cross--sections (described by $R_{2+1}$ in the denominator).  To
summarize the results of Fig. 2 we fitted $V(Q^2,y_c,r,s)$
by a linear function of $s$ and defined
\be
W(Q^2,y_c,r)\equiv \frac{\mr{d}V^{fit}(Q^2,y_c,r,s)}{\mr{d}s},
\label{W} \ee
which, by construction, is $s$--independent function of $Q^2,y_c$ and
$r$. Positive $W$ means that the variation of \lm in the PDF is more
important than that in the hard scattering cross--sections, while
for negative $W$ the situation is opposite.
The $Q^2$--dependence of $W(Q^2,y_c,r)$ is plotted in
Fig. 4a-c for three values of $y_c=0.01,0.02,0.04$ and five values of
$r=0.1,0.2,0.3,0.4,0.5$ GeV. In Fig. 5a we compare the $Q^2$ dependence
of $W(Q^2,y_c,r=0.2\;\mr{GeV})$ for different values of  
$y_c=0.01,0.02,0.04$. Similar plots can be drawn for other values of $r$ 
as well. From Figs. 2--5a we conclude that at the NLO:
\bit
\item For $Q^2$  below about $40$ GeV$^2$ the variation of $\lme$
is more important in PDF than in the hard scattering
cross-sections, while above $40$ GeV$^2$ the situation is reversed.
\item
Above $Q^2\approx 10^3$ GeV$^2$ the variation of \lm in PDF becomes
negligible.
\item The preceding conclusions depend only weakly on $y_c$.
\eit
At the LO the general features are the same as those displayed in Figs.
2--5a and we therefore merely summarize them in Fig. 5b. Comparing 
Figs. 5a and 5b we see that at the LO
\bit
\item the ``cross--over'' point $Q^2_{cr}$, where $W(Q^2_{cr},y_c,r)=0$,
is at higher values of $Q^2_{cr}\sim 50-200$ GeV$^2$ and depends more
sensitively on $y_c$, than at the NLO,
\item the relative importance of
varying \lm in PDF is for almost all values of $Q^2$ and $y_c$ larger
than at the NLO,
\item the difference between the LO and NLO results increases with 
increasing $y_c$.  
\eit

\subsection{The case of H1 cuts}
In \cite{H1} two subsamples of events with jets are defined:
\bd
\item {a)} high $Q^2$ sample: $Q^2 \ge 100$ GeV$^2$,
\item {b)} low $Q^2$ sample: $10\le Q^2\le100$ GeV$^2$.
\ed
In both subsamples outgoing jets angle in the HERA lab system was
restricted to lie in the interval 
$10^{\circ}\le\theta^{\mr{jet}}_{\mr{lab}} \le 145^{\circ}$, but there 
are also cuts on other variables, in which the two samples differ. In 
Fig. 6 plots analogous to those of Fig. 3, but incorporating the H1 
cuts, are displayed. We see that the experimental cuts suppress the 
sensitivity of $R_{2+1}(Q^2)$ to the variation of \lm in the PDF in the 
low $Q^2$ region, while leaving it essentially unchanged the high $Q^2$ 
one.  We conclude that in the region $Q^2\ge 100$ GeV$^2$, used in 
\cite{H1} for the extraction of \lm from the measured $R_{2+1}(Q^2)$, 
the variation of \lm in the PDF can be safely neglected.

\section{Conclusions}
In this paper we have shown that theoretically consistent determination
of $\alpha_s$ from $R_{2+1}$ in DIS requires the variation of $\lme$ in
both the hard scattering cross--sections and parton distribution
functions. After defining a procedure for the construction of PDF
which satisfy the corresponding evolution equations for an arbitrary
value of $\lme$, we used them to investigate the numerical importance of
varying $\lme$ in PDF for the quantity $R_{2+1}$. The results of this 
investigation show that if no cuts are applied on produced jets the 
variation of $\lme$ in PDF
\bit
\item is more important at the LO than at the NLO,
\item is more important than that in the hard scattering cross--sections
in low $Q^2$ region ($Q^2\le 40$ GeV$^2$ at the NLO and 
$Q^2\le 50-200$ GeV$^2$ at the LO),
\item can be safely neglected above $Q^2\approx 1000$ GeV$^2$.
\eit
We have also demonstrated that in the low $Q^2$ region the sensitivity
of $R_{2+1}(Q^2)$ to the variation of \lm in PDF is strongly suppressed 
by imposing experimental cuts applied in the H1 paper \cite{H1}.

\vspace*{0.5cm}
\noindent
{\Large \bf Acknowledgement} 

\noindent
This work had been supported in part by the grant No. A110136 of the 
Grant Agency of the Academy of Sciences of the Czech Republic.

\vspace*{0.6cm}
\noindent
{\Large \bf Appendix}

\vspace*{0.2cm}
\noindent
To show that \lm in PDF must be varied also in the ratio \reff{R2+1}
we shall draw on analogy between the cross--sections
$\sigma_{k+1}$ and the longitudinal and transverse structure functions
$F_{L}$, $F_2$ given as
\be
\frac{F_{2}(x,Q^2,\lme)}{x}=\int_{x}^{1}\frac{\mr{d}y}{y}
\left[\sum_{i=1}^{12}e_i^2q_i(y,M,\lme)
C_{q}^{(2)}(z,Q,M,\lme)
+\sum_{i=1}^{12}e^2_i
G(y,M,\lme)C_{G}^{(2)}(z,Q,M,\lme) \right], \label{F2}
\ee
\be
\frac{F_{L}(x,Q^2,\lme)}{x} = \int_{x}^{1}\frac{\mr{d}y}{y}
\left[\sum_{i=1}^{12}e^2_iq_i(y,M,\lme)
C_{q}^{(L)}(z,Q,M,\lme)
+\sum_{i=1}^{12}e^2_i G(y,M,\lme)C_{G}^{(L)}
(z,Q,M,\lme)
\right],
\label{FL}
\ee
where $z=x/y$, the sums run over all quark and antiquark flavours and
\bear
C_{q}^{(2)}(z,Q,M,\lme) & = &
\delta(1-z)+a(M/\lme)c^{(2)}_{q}(z,Q/M)+O(a^2(M/\lme)),
\label{c2q} \\
C_{G}^{(2)}(z,Q,M,\lme) & = & a(M/\lme)
c^{(2)}_{G}(z,Q/M)+O(a^2(M/\lme)), \label{c2G} \\
C_{q}^{(L)}(z,Q,M,\lme) & = & a(M/\lme)
c^{(L)}_{q}(z)+O(a^2(M/\lme)), \label{cLq} \\
C_{G}^{(L)}(z,Q,M,\lme) & = & a(M/\lme)
c^{(L)}_{G}(z)+O(a^2(M/\lme)). \label{cLG}
\eear
The quantity analogous to $R_{2+1}$ is the ratio
\be
R_L(x,Q^2,\lme)\equiv \frac{F_L(x,Q^2,\lme)}{F_2(x,Q^2,\lme)+
F_L(x,Q^2,\lme)}.
\label{Rx}
\ee
In terms of moments \reff{F2}-\reff{FL} read
\be
F_{2}(n-1,Q^2,\lme)=
\sum_{i}^{12}e_i^2q_i(n,M,\lme)C_{q}^{(2)}(n,Q,M,\lme)+
\sum_{i=1}^{12}e^2_i G(n,M,\lme)C_{G}^{(2)}(n,Q,M,\lme),
\label{F2n}
\ee
\be
F_{L}(n-1,Q^2,\lme)=
\sum_{i=1}^{12}e^2_i q_i(n,M,\lme)C_{q}^{(L)}(n,Q,M,\lme)+
\sum_{i=1}^{12}e^2_i G(n,M,\lme)C_{G}^{(L)}(n,Q,M,\lme).
\label{FLn}
\ee
For moments the quantity analogous to
$R_{2+1}(Q^2)$ is the ratio
\be
\overline{R}_{L}(n,Q^2,\lme)\equiv
\frac{F_{L}(n-1,Q^2,\lme)}{F_2(n-1,Q^2,\lme)+F_{L}(n-1,Q^2,\lme)}.
\label{RL}
\ee
Note that
$\overline{R}_L(n,Q^2,\lme)$ defined in \reff{RL} is not the moment
of the ratio $R_L(x)$, but rather the ratio of the moments!
We mention the analogy of $R_{2+1}$ to $R_L(x)$ and $\overline{R}_L(n)$
because all three quantities have similar structure, but the later two
are simpler and, moreover, $\overline{R}_L(n)$ can be treated
analytically. The technique of moments has recently been applied in
\cite{Hampel} to a related problem of extracting gluon distribution
function from jet production in DIS.

In the expressions for $\sigma_{k+1}, R_{2+1}$, as well as for
$F_{2,L}, R_L(x)$ and
$\overline{R}_L$, we have written out explicitly their
dependence on the value of $\lme$. Contrary to that, these observables
are formally (i.e. order by order of perturbation theory) independent of
the factorization scale $M$, the $M$--dependence of PDF being
cancelled by that of the hard scattering cross--sections.
In the case of $\overline{R}_L(n,Q^2,\lme)$, defined in \reff{RL},
the LO expression reads
\begin{eqnarray}
R_{L}(n,Q^2,\lme) &\!\! = \!\!& a(M/\lme)
 \frac{\sum_{i}e_i^2 q_i(n,M)c_{q}^{(L)}(n)+
 \left(\sum_{i}e^2_i\right) G(n,M,\lme)c_{G}^{(L)}(n)}
{\sum_{i}e^2_i q_i(n,M)} \label{R1} \nonumber \\
 &\!\! = \!\!& a(M/\lme) \left[c_q^{(L)}(n)+
 \frac{\left(\sum_{i}e^2_i\right) G(n,M,\lme)c_{G}^{(L)}(n)}
{\sum_{i}e^2_i q_i(n,M,\lme)}\right]+\cdots. \label{R2}
\label{RLLO}
\end{eqnarray}
We see that in the quark contribution to the LO term in \reff{RL}
quark distribution functions in the numerator and denominator of
\reff{RL} cancel out
\be
\overline{R}_L^{\mr{quarks}}=a(M/\lme)c_q^{(L)}+\cdots
\label{Rquarks}
\ee
and $\overline{R}_L^{\mr{quarks}}$ is thus equal to the parton
level hard scattering cross--section.
This, however, is not the case for the second term in \reff{RLLO},
corresponding to gluon contribution, which behaves rather like
$a^{1+\Delta(n)}$, where $\Delta(n)$ is, roughly speaking, the
difference between gluon and quark LO anomalous dimensions
\footnote{The situation is in fact even more complicated due to the fact
that quark and gluon distribution functions satisfy a system of coupled
evolution equations.}.
The cancellation of quark distribution functions in the LO coefficient
of \reff{RL} is a consequence of the fact that convolutions appearing in
\reff{F2} and \reff{FL} translate to simple multiplications for the
respective moments. The above mentioned cancellation holds only for
the ratio \reff{RL} of moments but not for the ratio \reff{Rx}
of structure functions themselves. This is obvious when we write
$F_{2,L}(x,Q^2)$ as expansions in orthogonal polynomials, which contain
infinite sums over different moments of PDF \cite{Barker,my}.
As a result, even in the quark contribution to $R_{2+1}$ variation of
\lm in PDF is in principle equally important as that in hard scattering
cross--sections. This
conclusion holds for the ratio $R_{2+1}(Q^2,\lme)$ as well.

\begin{figure}
\begin{center}
\epsfig{file=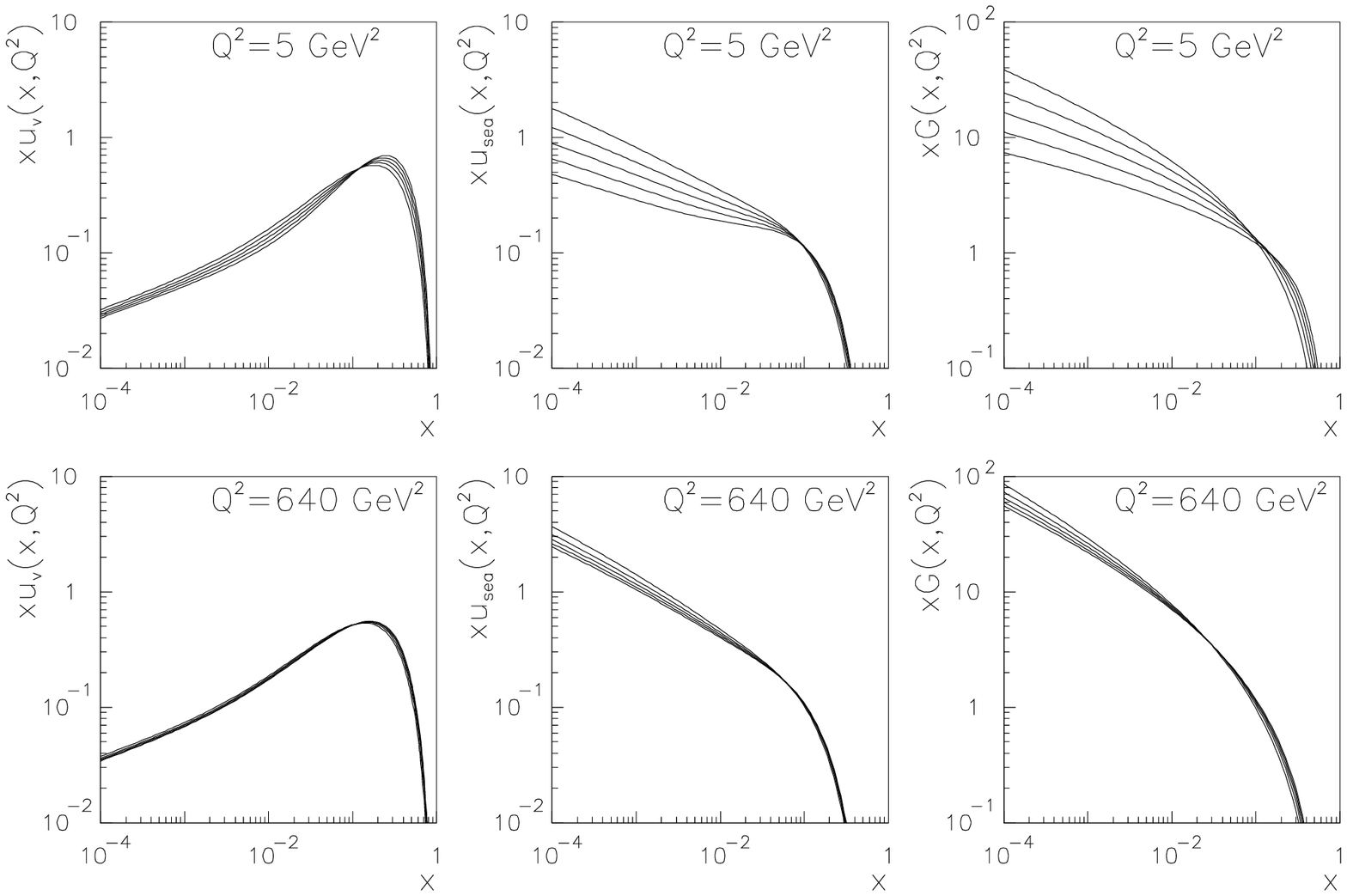,width=17cm}
\end{center}
\caption{The dependence of $xu_v(x,Q^2), xu_{\mr{sea}}(x,Q^2)$ and
$xG(x,Q^2)$ of the value $\lmepdf$ according to the recipe described
in the text and for two values of $Q^2$. At low $x$, the curves
correspond from above to the values of $\lmepdf$, subsequently,
$0.1,0.2,0.3,0.4$ and $0.5$.}
\end{figure}

\begin{figure}
\begin{center}
\epsfig{file=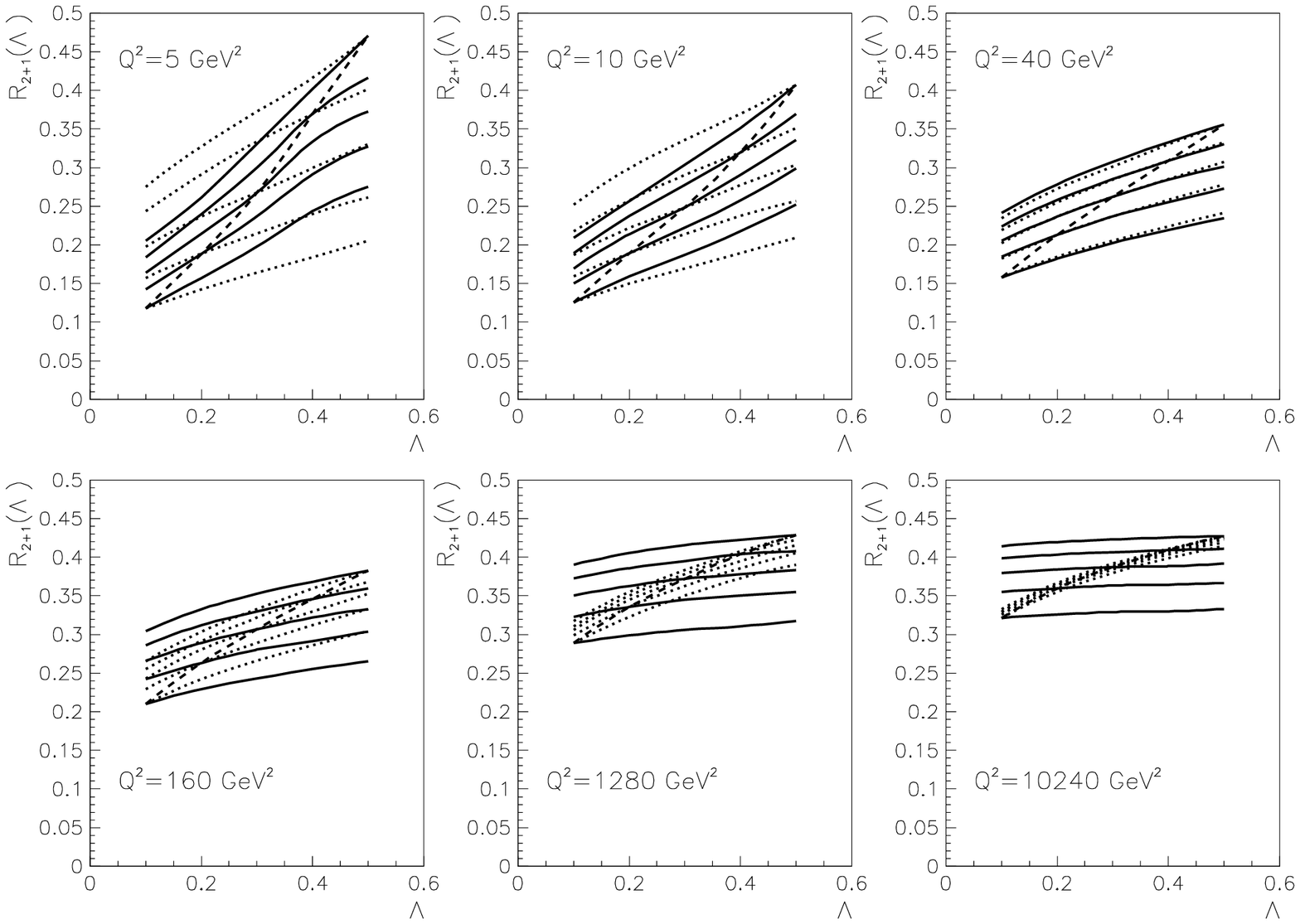,width=17cm}
\end{center}
\caption{The ratio $R_{2+1}(Q,y_c,\lmepdf,\lmehsc)$ as a function of
$\lmepdf$ for fixed $\lmehsc$ and a series of $Q^2$ values (solid
lines), or as a function of $\lmehsc$ for fixed $\lmepdf$ and the same
set of $Q^2$ values (dotted lines). The dashed lines correspond to
simultaneous variation of $\lmepdf=\lmehsc$. Solid curves
correspond, from below, to fixed $\lmehsc=0.1,0.2,0.3,0.4,0.5$ and the
dotted to the same fixed values of $\lmepdf$. In all plots $y_c$
equals $0.01$.}
\end{figure}

\begin{figure}
\begin{center}
\epsfig{file=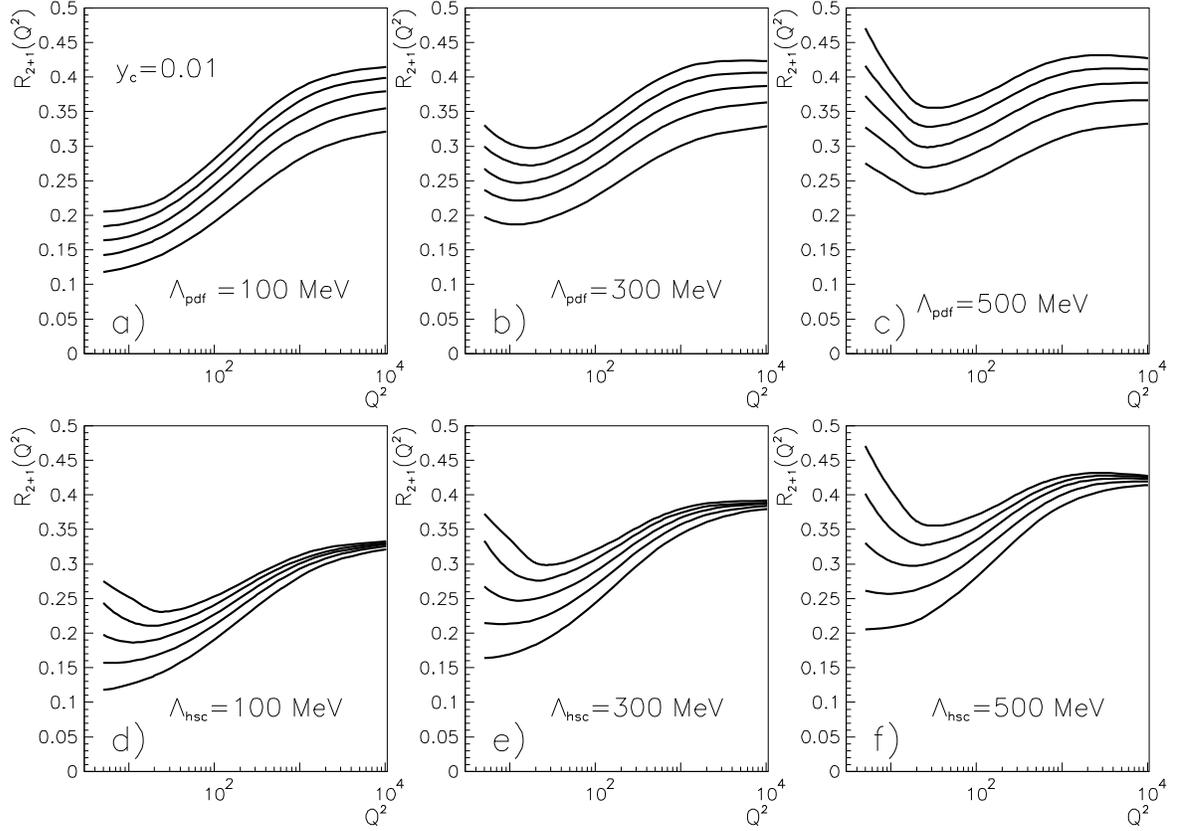,width=16cm}
\end{center}
\caption{(a--c) $Q^2$--dependence of the ratio $R_{2+1}(Q,y_c,\lmepdf,\lmehsc)$
for fixed $\lmepdf=0.1,0.3,0.5$ GeV and five values of
$\lmehsc=0.1,0.2,0.3,0.4,0,5$ GeV. (d--f) The role of
$\lmepdf$ and $\lmehsc$ is reversed. No cuts were applied in
evaluating $R_{2+1}$ and $y_c=0.01$. In all plots the curves
are ordered from below according to increasing $\lmehsc$ (in a--c) or
$\lmepdf$ (in d--f).} \end{figure}

\begin{figure}
\begin{center}
\epsfig{file=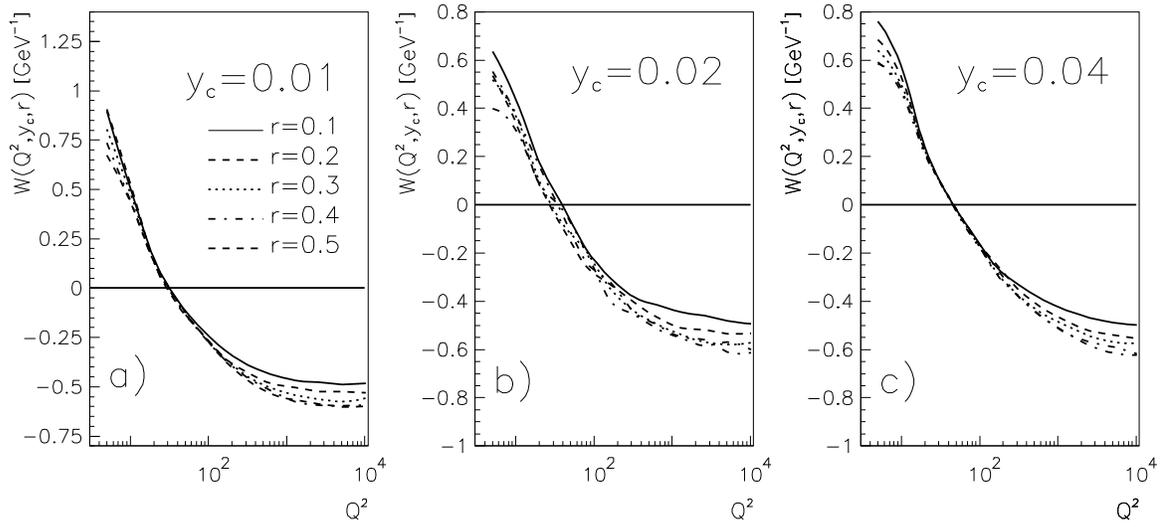,width=16cm}
\end{center}
\caption{(a--c) The quantity $W(Q^2,y_c,r)$ as a function of $Q^2$ for five
values of $r$ (given in GeV) and $y_c=0.01,0.02,0.04$. All curves
correspond to the case of no cuts.}
 \end{figure}

\begin{figure}
\begin{center}
\epsfig{file=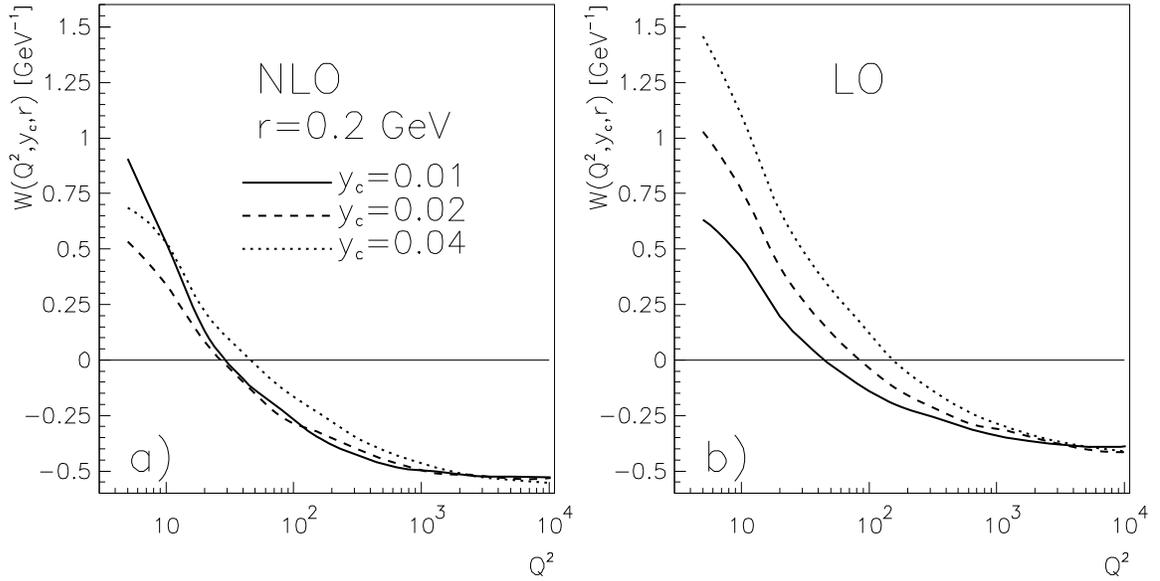,width=16cm}
\end{center}
\caption{The $Q^2$--dependence of $W(Q^2,y_c,r=0.2)$ for three
different values of $y_c$ at the NLO (a) and LO (b). All curves
correspond to the case of no cuts.}
\end{figure}

\begin{figure}
\begin{center}
\epsfig{file=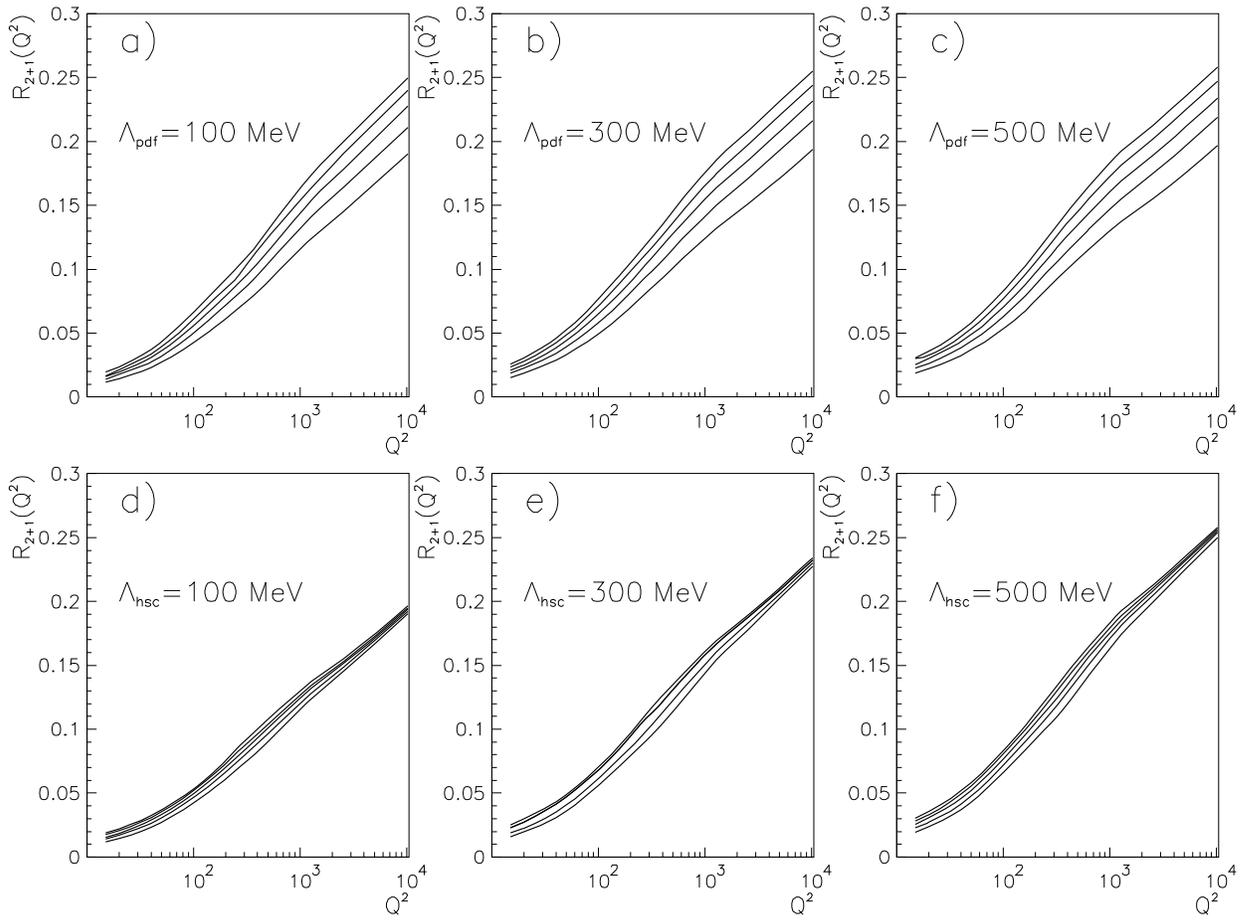,width=17cm}
\end{center}
\caption{The same as in Fig. 3 but for the H1 acceptance cuts.}
\end{figure}

\end{document}